# A Privacy-preserving Community-based P2P OSNs Using Broadcast Encryption Supporting Recommendation Mechanism

Ruihui Zhao[*], Mingjie Ding, Keiichi Koyanagi
Graduate School of IPS
Waseda University
Fukuoka, Japan
zachary@ruri.waseda.jp

Yuanliang Sun
School of information Science and Engineering
Southeast University
Nanjing, China

Liang Zhou
National Key Laboratory of Science and Technology on Communication
University of Electronic Science and Technology of China
Chengdu, China

*Abstract*—Online Social Networks (OSNs) have become one of the most important activities on the Internet, such as Facebook and Google+. However, security and privacy have become major concerns in existing C/S based OSNs. In this paper, we propose a novel scheme called a Privacy-preserving Community-based P2P OSNs Using Broadcast Encryption Supporting Recommendation Mechanism (PCBE) that supports cross-platform availability in stringent privacy requirements. For the first time, we introduce recommendation mechanism into a privacy-preserving P2P based OSNs, in which we firstly employ the Open Directory Project to generate user interest model. We firstly introduce broadcast encryption into P2P community-based social networks together with reputation mechanism to decrease the system overhead. We formulate the security requirements and design goals for privacy-preserving P2P based OSNs supporting recommendation mechanism. The RESTful web-services help to ensure cross-platform availability and transmission security. As a result, thorough security analysis and performance evaluation on experiments demonstrate that the PCBE scheme indeed accords with our proposed design goals.

## I. INTRODUCTION

Online Social Networks (OSNs) [1] are flourishing on the Internet, such as Facebook, Google+ and Twitter, all of which are centralized somehow. Centralized storage of personal data is a decisive factor for unintended information disclosure, because the existing OSNs providers may work on data mining, targeted advertising, and even information disclosure to third parties [2].

PeerSoN [3] have been proposed as the next generation OSNs based on P2P networks aiming to solving the privacy problems by getting rid of center provider (C/S model), however, it is restricted to only PeerSoN enabled devices. Guo et al [4] firstly combined P2P network structure with Web Services transmission structure, and they proved the feasibility of security transmission model. Unfortunately, they built secure communications by SOAP message which increased the transmission overhead largely and limited the expansion of networks. Our last paper ComPOSE [5] firstly combines Broadcast Encryption in community-based P2P OSNs, and realized similar functions in existing OSNs with low system cost. However, it does not support functions of recommending friends and groups.

In most existing OSNs, Facebook for example, recommendation mechanism [6][7] is an essential function because it deals with information overload by suggesting to users the groups and users that are potentially of their interests. The centralized server has all of users' personal information such as friend list, social graph, and users' interests, then it carries out data mining and existing recommendation systems to recommend friends and groups which accord with users' interests with high quality and accuracy. However, for a privacy-preserving P2P OSNs, how to efficiently and effectively transplant recommendation mechanism is still an open problem for two reasons [2]:

*1)* P2P OSNs cannot support powerful recommendation systems congenitally due to the loss of centralized server, who is responsible to collect users' information and compute similarities between users.

*2)* Users do not want to expose their friend list and interests to other nodes, including their own super nodes.

In this paper, we propose a Privacy-preserving Community-based P2P OSNs Using Broadcast Encryption Supporting Recommendation Mechanism scheme, in summary, this paper makes the following **CONTRIBUTIONS**:

- For the first time, we introduce recommendation mechanism into a privacy-preserving P2P OSNs. Besides, we firstly employ the Open Directory Project [8] to generate user interest model.
- Reputation mechanism in PCBE can make punishment on malicious nodes, which combines local trust and

global trust and successfully decreases the disk usage and network traffic.

- We take advantages of RESTful structure based on HTTP transmission, which makes our system available in cross- platform.

- Thorough security analysis and performance evaluations show that our scheme is security-enhanced and indeed accords with our design goals.

## II. DESIGN GOALS, SECURITY REQUIREMENTS

### A. Design Goals

In order to realize our PCBE scheme, the following security and performance guarantees should be simultaneously achieved:

- OSNs functions implementation: it can support OSNs functions such as AddFriends, StatusUpdates, OnlineChat, FileShare, and Cross-platform based on P2P structure with security enhancement supporting network scalability, high system efficiency.

- Privacy-preserving: to meet security requirements specified in Section II-B.

- Recommendation mechanism supporting: Given massive candidate users and groups, it can support efficient multi- keyword based similarities calculation with low over- heads and guarantee the most relevant users and groups to appear in the top-$k$ locations consistent with the target user's interest.

### B. Security Requirements

For the first time, we formally define the security requirements in privacy-preserving P2P OSNs supporting recommendation mechanism as follows:

- Data privacy: encryption algorithms should be implemented to ensure user profile privacy, storage security, and access control.

- Structure security: the integrity, applicability and active defense strategy of networks should be guaranteed.

- Transmission security: compatibility, attack resistance capability and transmission transparency among cross-platform are essential for conversations and data transmission. Messages should be standardized into unified formats, then protocols and encryption algorithms should be applied to maintain transmission security.

- Privacy-preserving recommendation mechanism: user's interest information and friendlist should be protected, which means indices and trapdoor should be constructed to prevent the super node from deducing the concealed interest keywords. In addition, PCEB should be able to prevent the super node from performing association attack, i.e., together with other malicious nodes, or deducing information of interest keywords from indices and trapdoor.

## III. THE DESIGN OF PCBE

### A. OSNs Functions Implement using Broadcast Encryption

We firstly proposed a privacy-preserving community-based P2P OSNs using broadcast encryption (PCBE) scheme, which is mainly based on P2P structured overlay networks [9].

All the peers in social networks would be divided into several groups or communities, every node has its interests and wants to join interested groups to consort with more friends. In Fig. 1, global network is consisted by super nodes, which are dynamic and selected by calculating reputation value which would be introduced in Section III-C, super nodes are similar to the role of servers in traditional C/S networks to some extent. For the sake of paper space, the details of broadcast encryption are not illustrated here, please refer to our previous paper called ComPOSE [5].

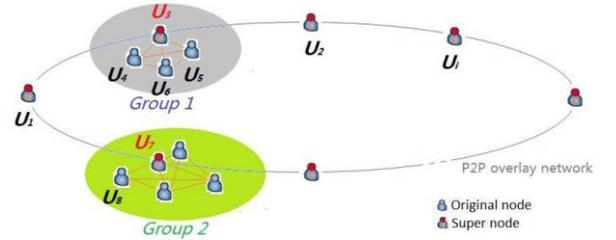

Fig. 1: Community-based broadcast encryption in PCBE

Our PCBE P2P-based OSNs provides users with various functions such as *adding friends, updating status, online chat*, and *file share*. They are implemented with one-to-one model and one-to-group model respectively. one-to-one model means that a single user contacts with another single user, they share a private key and start conversations in a private way afterwards, with which the functions of adding friends, online chat, and one-to-one file share are implemented. One-to-group model connects a single user to groups. Updating status and one-to-group file share functions are implemented with this model.

***One-to-one.*** Key exchange is applied in this model. *One-to-one* means one user contact another user through local network and then after confirmed by super node, messages would be sent in global network to other users in other groups.

***One-to-group.*** This model is quite similar to *one-to-one* model. The difference is that the terminal receiver is a group, or we could say all users in the group. Actually, we exchange messages between super nodes. For example, if $U_1$ wants to share his new photos with members in $G_2$, super node ($U_7$) of $G_2$ would calculate $Hash(G_2, U_1)$, and exchange this key with $U_1$, afterwards $U_1$ would share his photos with all the members in $G_2$. This model is quite useful when releasing notification messages, such as a conference notice or birthday party invitation to users in a group.

### B. Security Infrastructure

As illustrated in Fig. 2, in our PCBE OSNs, data privacy could be guaranteed by traditional cryptographic hash function (e.g., SHA1 [10]) to ensure data privacy, such as user interest profile, storage security, and access control.

As to the structure security, [11] has proved Broadcast Encryption could be applied in P2P overlay network based on communities to protect structure security. Besides, reputation mechanism is combined with Broadcast Encryption to guarantee the attack resistance capability of our proposed OSNs.

Transmission security is guaranteed by RESTful [12] web services based on security mechanism supporting cross-platform in our PCBE scheme. Our PCBE can employ flexible cryptology mechanism to protect data of different security levels, for example, basic data can simply be authenticated by *HMAC-SHA1*.

Privacy in recommendation mechanism is secured by the improved secure KNN proposed in our scheme and together with the reputation mechanism. We conducted a detailed security analysis in Section V.

Fig. 2: Security Infrastructure in PCBE

### C. Reputation mechanism

Reputation calculation is essential to protect structure security. It is someway an active defense strategy to make punishments on malicious peers. We directly adopt and simplify the reputation mechanism to calculate the trust scores of peers within the same community using local trust in [5] and within the different community using the global trust model in [13].

When a peer wants to send messages to other peers, it would calculate their reputation values at first, then it decides whether to send or not. In this way, malicious peers could be hard to get any messages or files in our network and finally be kicked out of networks. Due to the sake of the paper space, we did not show the details of reputation mechanism.

## IV. RECOMMENDATION MECHANISM

In this section, for the first time we propose a recommendation mechanism for the PCBE scheme using secure inner product computation, which is adapted and improved from a Secure KNN Computation [14], which is using the Euclidean distance, as the PCBE is using the inner product similarity of interest vectors instead of Euclidean distance, we need to do some modifications on the structure [8] [15] to adapt the PCBE framework. We will show how PCBE can recommend friends and groups that mostly accords with users' interest in a secure and privacy-preserving way with high privacy based on common interests as illustrated in Fig. 4, which mainly consists of the following four phases: *GenKey, BuildTrapdoor, BuildIndices*, and *CalculateScores*.

### A. Recommendation System Model

As illustrated in Fig. 3, our recommendation system model in PCBE scheme involves three different entities: the super node, the target user, and the candidate users and groups.

The procedure is summarized as follows: when the target user wants the system to recommend friends to him, he generates a trapdoor using the *BuildTrapdoor* step, and submit it to his super node. Afterwards he calculates the trust scores of nodes in his friend list, then sends *SK* to his trusted friends and asks them to send *SK* to their own trusted friends which is called candidate users here. Each candidate user $i$ build his index using *BuildIndices* step and submit it to the target user's super node. Then the super node calculates the scores for each Index using the trapdoor, ranks the scores and recommend top-$k$ candidate users to the target user.

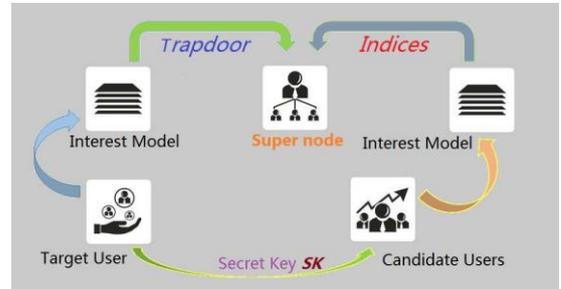

Fig. 3: Recommendation System Model

### B. Friends Recommendation based on common interests

*1) GenKey:* the target user randomly generates a $(n + 2)$-bit vector as $S$ and two $(n + 2) \times (n + 2)$ invertible matrices *{$M_1$, $M_2$}*, where $n$ denotes the number of keywords in the keyword dictionary. The secret key *SK* is in the form of a 3-tuple as *{S, $M_1$, $M_2$}*.

*2) BuildTrapdoor($\widetilde{W}$, SK):* the target user donates his interest keywords as $\widetilde{W}$ according to his interest model, then an $n$-bit binary vector $q$ is generated where each bit $q[j]$ indicates whether $W_j \in \widetilde{W}$ is true or false, $j \in [1, n]$.

In vector $q$, for each bit whose value is equal to *1*, the target user searches in his interest model for its corresponding preference weight donated as $p_i$, $i \in [1, m]$ and substitutes the mentioned value *1* in the vector q, which is denoted as $Q$, the left entries in $Q$ are set to *0*. $Q$ is firstly extended to $(n+1)$-dimension which is set to *1*, and then scaled by a random number $r$ not equal to *0*, and finally extended to a $(n+2)$-dimension vector as $\vec{Q}$ where the last dimension is set to another random number $t$. $\vec{Q}$ is therefore equal to *(rQ, r, t)*.

For $j = 1$ to $(n + 2)$, if $\vec{S}[j] = 0$, $\vec{Q'}[j]$ and $\vec{Q''}[j]$ are set to two random numbers so that their sum is equal to $\vec{Q}[j]$; else, $\vec{Q'}[j]$ and $\vec{Q''}[j]$ are set the same as $\vec{Q}[j]$. Finally, the trapdoor $T_{\widetilde{W}} = \{M_1^{-1}\vec{Q'}, M_2^{-1}\vec{Q''}\}$ is built for the target user's interest vector $\vec{Q}$. Finally, he sends $T_{\widetilde{W}}$ to the super node whose trust score is highest among the SuperNodes of the groups he joined in according to the reputation mechanism mentioned above.

***Interest model generation algorithm:*** the Open Directory Project [8] has a huge directory structure, the lower sub class is some different classification of the upper class from different perspectives, thus we only reserve the branches that users need: our PCBE scheme supplies detailed classification directory for the user to choose and generates the initial interest keywords when the user registers the PCBE OSNs for the first time. Afterwards the model extracts keyword sets from the interest keywords of groups which the user joins in. At the time when a super node construct a group, he chooses interest keywords for this group according to the Open Directory Project, once nodes join in this group, the super node will send them the interest keywords profile of this group. In conclusion, the preference generation algorithm is shown as *Algorithm 1*.

---
**Algorithm 1** Preference generation algorithm of user interest model
**Input:** User interest keywords set X, Group interest keywords sets Y, keyword vector $y_i$ in Y
**Output:** Updated user interest keywords set X'
1: **while** ($y_i \neq \emptyset$) **do**
2:    bring keyword $k_{i,l}$ from $y_i$
3:    **for** (each keyword $k_{i,l}$ in $y_i$) **do**
4:      **if** ($k_{i,l}$ exists in $x_j$) **then**
5:         $p_{i,l} = p_{i,l} + 1$
6:      **else**
7:         create new node for keyword $k_{i,l}$ in $x_j$
8:         $p_{i,l} = 1$
9:      **end if**
10:    **end for**
11:    delete $y_i$ from Y
12: **end while**
13: **return** X'

---

*3) BuildIndices($D_i$, SK):* the target user calculate the trust scores of nodes in his friend list, then send *SK* to his trusted friends and ask them to send them to their own trusted friends which is called candidate users in this paper.

Candidate user *i* generates a *n*-bit interest vector $D_i$ according to his interest model. When $j \in [1,n]$, each bit $D_i[j]$ denotes the preference weight value of the keyword, if $D_i[j]= 0$, it means the keyword in this position does not appear in his interest model. Extend $D_i$ to *(n+2)*-bit, where the *(n+1)*-th entry is set to a random number $\varepsilon_i$, and the (n+2)-th entry is set to *1*. $\vec{D_i}$ is therefore equal to $(D_i, \varepsilon_i, 1)$.

Every plaintext index $\vec{D_i}$ is then split into data vector pair denoted as $\{\vec{D_{i'}}, \vec{D_{i''}}\}$ using the splitting process of the secure *k*-nearest neighbor (kNN) scheme [14] as follows: for $j = 1$ to $(n+2)$, if $\vec{S}[j] = 1$, then $\vec{D_{i'}}[j]$ and $\vec{D_{i''}}[j]$ are set to two random numbers so that their sum is equal to $\vec{D_i}[j]$; else, $\vec{D_{i'}}[j]$ and $\vec{D_{i''}}[j]$ are set as the same as $\vec{D_i}[j]$. Finally, the index $I_i = \{M_1^T \vec{D_{i'}}, M_2^T \vec{D_{i''}}\}$ is built for every candidate user. Finally, he sends $I_i$ to the target user's specified super node.

*4) CalculateScores($T_{\widetilde{W}}$, k, I):* for each candidate user, with the trapdoor $T_{\widetilde{W}}$, the super node computes the similarity scores as shown in the following equation, ranks all scores and returns the top-*k* ranked candidate users' identification to the target user, then he sends an request of adding friends to them.

$$T_{\widetilde{W}} \cdot I_i = \{M_1^{-1}\vec{Q'}, M_2^{-1}\vec{Q''}\} \cdot \{M_1^T\vec{D_{i'}}, M_2^T\vec{D_{i''}}\} \quad (1)$$
$$= \vec{Q'} \cdot \vec{D_{i'}} + \vec{D_{i''}} \cdot \vec{Q''} = \vec{Q} \cdot \vec{D_i}$$
$$= (rQ, r, t) \cdot (D_i, \varepsilon_i, 1)$$
$$= r(Q \cdot D_i + \varepsilon_i) + t$$
$$= r\underbrace{\sum_{i=1}^{m}(p_i \cdot H_{x_i})}_{x_i \in [1,n]} + r\varepsilon_i + t$$

*C. Groups Recommendation*

Groups Recommendation is almost the same as the procedure of recommending friends, except in the step of generating indices: after the target user executes the steps of *GenKey*, *BuildTrapdoor*, he calculates the trust scores of nodes in his friend list, then send SK to his trusted friends and ask them to send them to their own joined groups, namely their corresponding super nodes.

If the target user's specified super node is in the list, then delete this super node, for it means this group has been in the group list of target user. Then super nodes carry out the *BuildIndices* step, using the corresponding interest keywords profile of this group. The rest procedures are the same as listed in Section IV-B.

## V. SECURITY ANALYSIS

In this section, we analyze the security properties under the schemes we introduced above. We will focus on four aspects: data privacy, structure security, transmission privacy, privacy in recommendation mechanism, which demonstrates that the PCBE scheme indeed accords with our proposed Security Requirements.

*A. Data privacy*

Traditional cryptographic hash function (e.g., SHA-1 [10]) could be properly utilized here to guarantee data privacy such as user interest profile. Although SHA-1 is no longer sufficient in some scenarios, it could be replaced by other existing encryption techniques, and it not within the scope of this paper.

*B. Structure security*

[11] has proved Broadcast Encryption could be applied in P2P overlay network based on communities to protect structure security. Besides, reputation mechanism in our scheme helps guarantee the attack resistance capability of our proposed OSNs.

*C. Transmission privacy*

RESTful web services based on security mechanism is proposed to extend secure transmission in cross-platform. In this paper, we determine to implement Oauth [16] for RESTful. Besides, our PCBE can employ hierarchy security mechanism flexibly, for example, basic data can be authenticated by HMAC-SHA1. For Middle-level data, PCBE can employ random string in message envelop. For high-level data such as password recovery and private conversations, dynamic password can be added after request parameters which is proved suitable and secure in [17].

*D. Privacy in recommendation mechanism*

The indices and trapdoor are well protected if the secret key SK is kept confidential since such vector encryption method has been proved to be secure in [14]. With the randomness introduced by the splitting process and the random numbers $r$ and $t$, our PCBE scheme can generate two totally different trapdoors and indices even for the same interest vector. This nondeterministic generation algorithm can guarantee the unlinkability of any two trapdoors or indices. Moreover, with properly selected parameter $\sigma$ for the random factor $\varepsilon_i$, even the final score results can be obfuscated very well, preventing the super node from learning the relationships of given trapdoors or indices and the corresponding interest keywords.

Besides, as shown in the recommendation system model, indices are sent from candidate users who are not in the user's friendlist, so super node cannot deduce the friendlist of the target user. Our reputation mechanism can help prevent the super node from performing association attack together with other malicious nodes, for *SK* are sent to the target user's trusted friends, which can guarantee the candidate users will not expose *SK* to the specified super node.

## VI. EVALUATION

### A. System Performance

For overall evaluation, we compare our proposal PCBE with Safebook, Peerson, etc. Fig. 4 shows that our scheme performs better in supporting OSNs functions, strong authentication, reputation mechanism, network scalability, cross-platform compatibility and recommendation systems, which demonstrates that the PCBE scheme indeed accords with our proposed design goals.

|  | Safebook | Peerson | Batch-Authenticated | PCBE |
|---|---|---|---|---|
| OSNs Functions Supporting | ○ | ○ | √ | √ |
| Strong Authentication | √ | √ | √ | √ |
| Reputation Mechanism | √ | × | ○ | √ |
| Network Scalability | MIDDLE | MIDDLE | HIGH | HIGH |
| Cross-platform Compatibility | × | ○ | ○ | √ |
| Recommendation Mechanism | × | × | × | √ |

√ Supported  × Not Supported  ○ Partly Supported

Fig. 4: System performance comparisons of PCBE and other existing systems

### B. Broadcast Encryption

We evaluate system efficiency to prove our scheme performs better by properly choosing encryption algorithms, which is derived and simplified from our last paper [5]. Based on communication cost, we figure out the overall system average cost in Fig. 5. We set user scale increasing from *1* to *80* million to simulate the real situations, and make *50* cycles to make the result accurate.

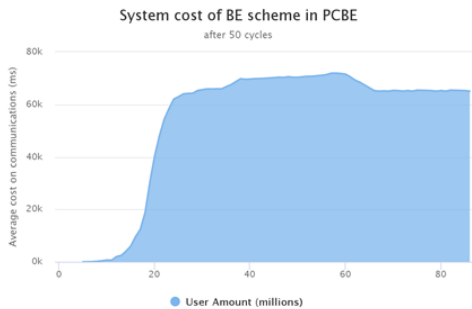

Fig. 5: System efficiency of PCBE

As shown in Fig. 5, when user amount increases from *0* to *20* million, system cost of PCBE keeps increasing very fast up to about *65* seconds, which is mainly caused by the huge number of broadcast headers. When user amount is increasing from *20* million to *60* million, the trend seems not increasing so quickly, which means as users are joining in groups and forming communities, our communication cost could be under control. What we should note here is when user amount passes over *60* million, using BE scheme would obviously decrease the system cost due to the community-based broadcast encryption structure, because at this moment most of users have joined in their favorite groups. In addition, at this period our system cost stays smooth which also indicates our network performs stable even if user amount is increasing.

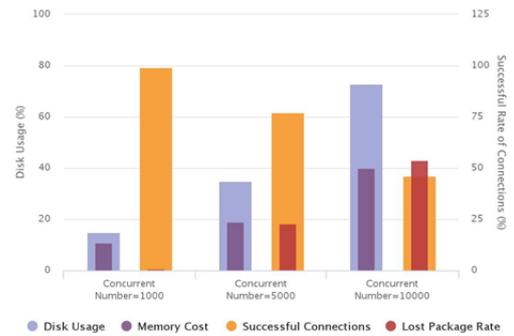

Fig. 6: The disk usage, memory cost, etc. with concurrent request number = 1000, 5000, 10000 respectively

Besides, system performance of P2P OSNs with web services could be analyzed by concurrent connections. In this paper, connections including RESTful will be influenced by network state and network facilities. A successful connection needs a working thread for executing request codes and a completion port thread for receiving recall from RESTful servers. Therefore, if multiple RESTful calls occur at the same time, the thread pool would be run out. As we can see from Fig. *6*, our successful connection rate would reduce as concurrent number is increasing, and stay at almost *50%* when concurrent number is *10,000*. This scale of concurrent connections in OSNs would be enough to handle most system requirements, because a successful connection rate of *50%* would lead to a result that the success rate of each command can reach larger than *95%* [4], which ensures that important requests could be transmitted stably. However, the disadvantage is that disk Usage and memory cost would both increase as concurrent number is increasing. For example, when concurrent number is *10,000*, lost package rate would reach about *40%*.

### C. Recommendation Mechanism

We refer to the performance evaluation methods in [8] and [15]. As far as we know, PCBE is the first scheme to support recommendation mechanism in P2P OSNs in a privacy-preserving way. We did not find other existing work in the literature to compare with from the view of recommendation quality and system cost.

*1) Recommendation Quality:* assume the true score of calculating the similarities between the target user and a

candidate user is $x_i = Q \cdot D_i$. From the calculate scores equation, the final similarity score as $y_i = r(Q \cdot D_i + \varepsilon_i) + t = r(x_i + \varepsilon_i) + t$ is a linear function of $x_i$, where the coefficient $r$ is set as a positive random number. However, because the random factor $\varepsilon_i$ is introduced as a part of the similarity score, the final search result on the basis of sorting similarity scores may not be as accurate as that in plaintext similarity calculation. For the consideration of accuracy, we can let $\varepsilon_i$ follow a normal distribution $N(\mu, \sigma^2)$, where the standard deviation σ functions as a flexible trade-off parameter among similarity calculation accuracy and security. From the consideration of recommendation quality, $\sigma$ is expected to be smaller so as to obtain high precision indicating the good quality of recommendation results.

*2) Recommendation Efficiency:*

### A. Generating Indices/Trapdoor

As presented in Section IV-B, the major computation of building indices and trapdoor includes the splitting process and two multiplications of a $(n + 2) \times (n + 2)$ matrix and a $(n + 2)$-dimension vector. As shown in Fig. 7, the time of generating a trapdoor or index is greatly affected by the number of keywords in the dictionary, here $k$ is the number of interest keywords. Fig. 8 describes that the number of keywords in the interest model has little influence upon the result because the dimensionality of vector and matrices is always fixed with the same dictionary.

As shown in Tab. I, we compare the storage overhead of indices, trapdoor in PCBE within different sizes of dictionary. The size is absolutely linear with the size of dictionary.

In addition, Tab. II lists the storage overhead of the user profile model with different numbers of keywords which is denoted as *m*, we can draw a conclusion that it is absolutely linear with the value of *m* and is acceptably negligible.

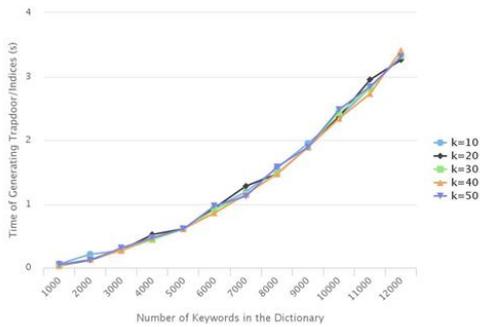

Fig. 7: For different numbers of keywords in the dictionary.

### B. Calculating Scores

The major computation to calculate scores for the super node consists of computing and ranking similarity scores for candidate users in the candidate space and selecting top-*k* results from all the scored candidate users. Fig. *9* shows the results of time consumption of calculating scores for different numbers of keywords within the same candidate space of *10,000* candidate users, Fig. *10* shows the time cost of calculating scores for different numbers of candidate users within the same dictionary, we set *k* to *50* in our experiments. We can learn that the time of calculating scores is linear with both the number of users in the candidate space and the size of dictionary.

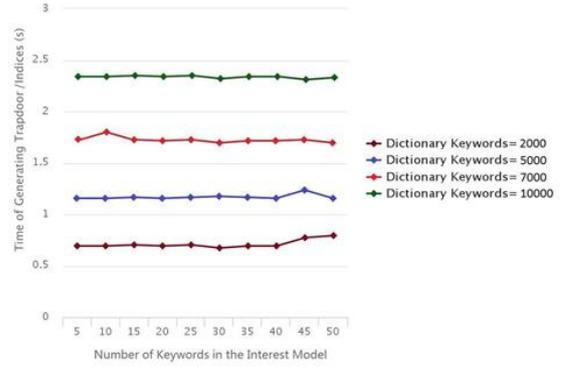

Fig. 8: For different numbers of keywords in users' interest model within the same dictionary of keywords.

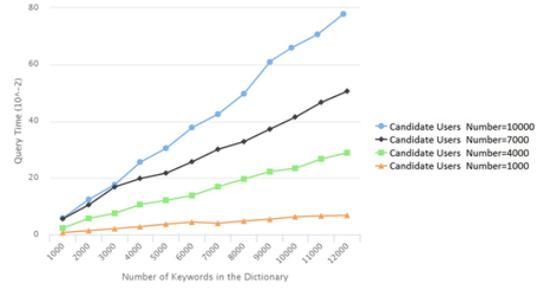

Fig. 9: Time of calculating scores for different numbers of keywords within the same candidate space

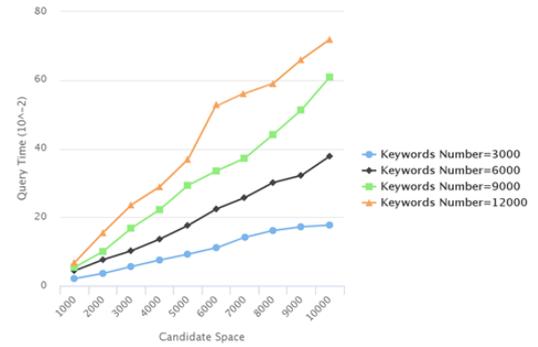

Fig. 10: Time of calculating scores for different numbers of candidate users.

TABLE I. SIZE OF SUBINDEX/TRAPDOOR

| Dic_size | 4000 | 6000 | 8000 | 10000 | 12000 |
|---|---|---|---|---|---|
| Trapdoor(KB) | 31.2656 | 46.8906 | 62.5156 | 78.1406 | 93.7500 |
| Indices (KB) | 31.2656 | 46.8906 | 62.5156 | 78.1406 | 93.7500 |

TABLE II. SIZE OF USER INTEREST PROFILE

| m | 50 | 100 | 200 | 400 | 800 |
|---|---|---|---|---|---|
| PCBE (KB) | 0.3906 | 0.7813 | 1.5625 | 3.1250 | 6.2500 |

## VII. CONCLUSION

In this paper, we propose a Privacy-preserving Community-based P2P OSNs Using Broadcast Encryption Supporting Recommendation Mechanism called PCBE. We firstly implement community-based broadcast encryption in P2P OSNs supporting recommendation mechanism with OSNs functions feasibility and low system cost reasonably. RESTful web services based on flexible security mechanism is proposed to guarantee the transmission security and support cross-platform with low overhead. Thorough Security analysis shows that the proposed scheme accords with our formulated privacy requirements. Extensive experiments demonstrate that PCBE achieves the design goals. In the future, we will further investigate how PCBE collaboratively collects online social graph information in a privacy-preserving way.